\documentclass[a4paper]{article}
\usepackage[utf8]{inputenc}
\usepackage{cochineal}
\usepackage[cochineal]{newtxmath}
\usepackage{roboto}
\usepackage[cal=boondoxo]{mathalfa}
\usepackage{amsmath,graphicx,microtype,bm,xcolor,siunitx}
\usepackage[numbers,sort&compress]{natbib}
\definecolor{boxcolor}{cmyk}{1, 0.5, 0, 0.1}
\definecolor{linkscolor}{cmyk}{0.6, 0.3, 0, 0.9}

\usepackage[colorlinks=true,allcolors=linkscolor!60]{hyperref}

\usepackage{titlesec}
\usepackage{titling}
\titleformat*{\section}{\Large \bfseries \sffamily}
\titleformat*{\subsection}{\large \bfseries \sffamily}
\titleformat*{\subsubsection}{\bfseries \sffamily}

\usepackage[centering,margin=3.5cm]{geometry}

\usepackage{tcolorbox}
\newtcolorbox{abstractbox}{
  arc=0pt,
  boxrule=0pt,
  colback=boxcolor!20,
  boxsep=1em,
  left=0pt, right=0pt, bottom=0pt, top=0pt,
  width=1\columnwidth
}

\renewenvironment{abstract}{
   \noindent
   \flushleft
   \begin{minipage}{0.9\columnwidth}
   \upshape\sffamily 
   \begin{abstractbox}
   \fontsize{9}{14}\selectfont
  }{
   \end{abstractbox}
   \end{minipage} 
   \vskip 2.0em
  }

\usepackage[font={small}]{caption}

\makeatletter
 \def\@textbottom{\vskip \z@ \@plus 1pt}
 \let\@texttop\relax
 \def\NAT@def@citea{\def\@citea{\NAT@separator\,}}
 \let\@fnsymbol\@arabic
\makeatother

\pretitle{\begin{flushleft}\LARGE \bfseries}
\posttitle{\par\end{flushleft}\vskip 0.5em}
\preauthor{\begin{flushleft}
\large \lineskip 0.5em%
\begin{tabular}[t]{c}}
\postauthor{\end{tabular}\par\end{flushleft}}
\predate{\begin{flushleft}\footnotesize \itshape}
\postdate{\par\end{flushleft}}

\title{Pair spin-orbit interaction in low-dimensional electron systems}

\author{Yasha Gindikin\thanks{e-mail: \href{mailto:gindikin@protonmail.ch}{gindikin@protonmail.ch}}\; and Vladimir A.\ Sablikov\thanks{e-mail: \href{mailto:sablikov@gmail.com}{sablikov@gmail.com}}
}
\date{Kotelnikov Institute of Radio Engineering and Electronics, Russian Academy of Sciences, Fryazino, 141190, Russia}

\begin{document}
\maketitle

\begin{abstract}  
The pair spin-orbit interaction (PSOI) is the spin-orbit component of the electron-electron interaction that originates from the Coulomb fields of the electrons. This relativistic component, which has been commonly assumed small in the low-energy approximation, appears large and very significant in materials with the strong SOI\@. The PSOI, being determined by the spins and momenta of electrons, has highly unusual properties among which of most interest is the mutual attraction of the electrons in certain spin configurations. We review the nature of the PSOI in solids and its manifestations in low-dimensional systems that have been studied to date. The specific results depend on the configuration of the Coulomb fields in a particular structure. The main actual structures are considered: one-dimensional quantum wires and two-dimensional layers, both suspended and placed in various dielectric media, as well as in the presence of a metallic gate. We discuss the possible types of the two-electron bound states, the conditions of their formation, their spectra together with the spin and orbital structure. In a many-particle system, the PSOI breaks the spin-charge separation as a result of which spin and charge degrees of freedom are mixed in the collective excitations. At sufficiently strong PSOI, one of the collective modes softens. This signals of the instability, which eventually leads to the reconstruction of the homogeneous state of the system.
\end{abstract}

\section{Introduction}

It is known from the relativistic quantum mechanics that the interaction between charged particles depends not only on the their charge and mutual distance but also on their spins and momenta~\cite{PhysRev.34.553,PhysRev.36.383,PhysRev.39.616}. This interesting fact has not been generally appreciated in condensed matter physics. However recent studies~\cite{PhysRevB.95.045138,doi:10.1002/pssr.201700256,2017arXiv170700316G,2018arXiv180410826G,PhysRevB.98.115137,10.1016/j.physe.2018.12.028,2019arXiv190409510G} have found out that it is very important in materials with the strong Rashba spin-orbit interaction (SOI). In this case, the spin-orbit component of the pair interaction Hamiltonian, which is small in the non-relativistic approximation in vacuum~\cite{bethe2012quantum}, becomes very large similarly to the Rashba SOI~\cite{winkler} and gives rise to a significant reconstruction of the correlated electron state. To distinguish this effect from the Rashba SOI we call the spin-orbit component of the electron-electron (e-e) interaction the pair spin-orbit interaction (PSOI). The main feature of the PSOI is that it essentially depends on the spins and momenta of the interacting particles, not to mention the more complex spatial dependence of the interaction strength.

In this paper we review the origination of the strong PSOI in crystals, most interesting modifications of the PSOI in actual low-dimensional structures caused by specific configuration of the Coulomb fields therein, and basic effects due to the PSOI that have been studied to date.

The key aspect of the PSOI is that it creates an attraction between the electrons in certain spin configurations tied to their momenta~\cite{PhysRevB.95.045138}. The e-e attraction emerges from the well-known fact that the larger the electric field creating the SOI, the lower the bottom of the conduction band. When the electrons approach each other, the increased pair Coulomb field shifts the conduction band downwards to lower the energy for an electron with a particular spin orientation, which exactly means the e-e attraction.

The first remarkable effect of the attractive PSOI is the electron pairing, which occurs for sufficiently strong PSOI attainable in modern materials~\cite{2018arXiv180410826G,PhysRevB.98.115137,10.1016/j.physe.2018.12.028,2019arXiv190409510G}. Depending on the character of the electron motion that creates the PSOI, there arise two kinds of two-electron bound states. The \emph{relative} bound states are formed by the motion of electrons with respect to each other. On the contrary, the motion of the electron pair as a whole forms the \emph{convective} bound states, the binding energy of which crucially depends on the total momentum of the pair. The binding energies are estimated to be in the meV range in modern materials with giant SOI, and can be tuned by the gate voltage.

The PSOI has even more interesting manifestations in the many-particle systems. To date, they have been investigated for the strongly correlated electrons in 1D quantum wires~\cite{PhysRevB.95.045138}, where the Luttinger liquid is formed as a result of the e-e interaction. The cornerstone of the Luttinger liquid is the spin-charge separation, which manifests in the existence of plasmons and spinons~\cite{voit1995one}. The PSOI breaks the spin-charge separation, because of which the spin and charge degrees of freedom are mixed in the collective excitations. The PSOI signatures in the spin-charge structure of the collective excitations and in their renormalized velocities can be quite simply identified by the Fabry-Pérot resonances in the frequency-dependent conductance of the quantum wire coupled to leads~\cite{2017arXiv170700316G}.

Notably, the PSOI leads to a strong softening of one of the collective modes in the long-wave region that evolves from a pure spin excitation to a pure charge one. In other words, the 1D electron liquid becomes unstable with respect to the long-wave fluctuations of the electron density. At the instability threshold, the mode velocity turns to zero together with the charge stiffness of the system~\cite{PhysRevB.95.045138}. 

The plan of the review is as follows. In Section~\ref{sec2}, we outline how the PSOI Hamiltonian in crystalline solids can be formulated on the basis of the Breit-Pauli Hamiltonian within the $k \cdot p$ approximation, and discuss under which conditions and in which materials the PSOI contribution to the e-e interaction becomes significant. The configuration of electric fields leading to specific forms of the PSOI Hamiltonians is discussed in Section~\ref{elstat} for different low-dimensional structures, whereas an overview of the electron pairing theory is given in Section~\ref{pairing}. Section~\ref{transport} is focused on the transport manifestations of the PSOI in quantum wires.

\section{Pair spin-orbit interaction in solids}
\label{sec2}

Electrons in solids form a fruitful ground to raise many ideas born in relativistic quantum theory and to give life to plenty of unexpected phenomena arising from the realization of electron states and band spectra that only recently appeared truly exotic. This became possible largely due to the discovery of new materials and technological advances in nanostructures.

Just recall the discovery of graphene~\cite{geim2009rise} and carbon nanotubes~\cite{dresselhaus2000carbon}, topological insulators~\cite{valenzuela2015topological,RevModPhys.88.021004,shun2017topological}, Weyl and Dirac semi-metals~\cite{RevModPhys.90.015001}, 2D transition metal dichalcogenides~\cite{manzeli20172d}, and many more.

The wide variety of non-trivial electronic states and spectra is due to the presence of the crystal. Electron motion in the crystal potential is generally speaking described by the relativistic Dirac equation~\cite{dirac1928quantum}, but in practice a quasi-relativistic approximation based on a small ratio $v/c$ of the electron velocity to the speed of light is sufficient to describe the electronic spectrum of any material. Such an approximation is successful, in particular, in describing the behavior of electron spins, which has opened a whole new land of spin phenomena in solids~\cite{Dyakonov2017}. 

Historically, the first quasi-relativistic approximation was developed by Pauli~\cite{pauli1980}, who derived relativistic corrections to the Schrödinger equation that include the SOI term
\begin{equation}
\label{pauliso}
    H_\mathrm{SOI} = \frac{e \hbar}{4 m^2 c^2} \left(\mathbf{E}\times \mathbf{p}\right) \cdot \bm{\sigma} \,,
\end{equation}
as well as the correction to the kinetic energy (the mass-velocity operator), and Darwin operator.  In Eq.~\eqref{pauliso}, $\mathbf{E}$ is the electric field acting on the electron, $\mathbf{p} = -i \hbar \nabla$ is the momentum of electron, and $\bm{\sigma}$ is the Pauli vector. The SOI has attracted a great deal of attention in solid state physics, as it was found that the coupling of the electron spin to its orbital motion in crystals resulted in a dramatic reconstruction of the electron band spectra and states~\cite{PhysRev.100.580,Rashba.Sheka}. 

SOI manifestations in crystalline solids are investigated both in electron spectra~\cite{bir1974symmetry} and, which is most important for us here, in the electron dynamics~\cite{glazov2018electron}. The dynamics of electrons is described in terms of the envelope wave-functions using the $k \cdot p$ method~\cite{voon2009kp}. This approach leads to the equations of motion similar to the Pauli-Schrödinger equation, but with material-dependent SOI constants.

\subsection{One-body spin-orbit interaction}

In low-dimensional systems there exist two quite distinct types of the SOI, depending on the source of the electric field $\mathbf{E}$ that produces the SOI\@. SOI produced by the crystalline field in crystals with bulk inversion asymmetry (BIA) is called the BIA SOI\@. Also, SOI can be produced by the electric fields external to the crystal. This is commonly called the Rashba SOI~\cite{bychkov1984properties,winkler,bihlmayer2015focus,manchon2015new}. In quantum structures, the common source of this field is the confining potential if its profile creates a structure inversion asymmetry (SIA). 

Let us emphasize that the Rashba SOI can be created by other sources, too. The most important examples of such sources are charged impurities and structure defects~\cite{SMIT195839,Perel}. If the electric field created by the defect is variating smoothly on the scale of the lattice constant, the SOI produced by this field can be described similar to the Rashba SOI even in the absence of the SIA~\cite{PhysRevB.2.4559,Nozieres,PhysRevB.64.014416,PhysRevLett.95.166605}. This is the basis of the theoretical description of the extrinsic spin-Hall effect~\cite{RevModPhys.87.1213}. 
 
The Rashba SOI Hamiltonian has the form
\begin{equation}
\label{rham}
        H_{\mathrm{RSOI}} = \frac{\alpha}{\hbar} (\mathbf{E} \times \mathbf{p})\cdot \bm{\sigma}\,,
\end{equation}
where $\alpha$ is a material-dependent SOI constant that exceeds a SOI constant in vacuum by a huge factor, the specific value of which depends on the crystalline potential and the properties of the electron band states.

\subsection{Two-body spin-orbit interaction}

The SOI manifestation in low-dimensional structures is not exhausted by the simple single-electron picture presented above  because it ignores the e-e interaction, which can completely change the single-electron description~\cite{giamarchi2003quantum,martin2016interacting}.

The very first problem one encounters when describing the interacting electron system is to derive the e-e interaction Hamiltonian. Most often, one simply adopts the Coulomb interaction potential. We propose to treat the problem consistently, similar to the derivation of the single-particle Hamiltonian in the crystal from the Dirac equation.   

The theory of pair interaction in relativistic quantum mechanics was developed by Breit~\cite{PhysRev.34.553,PhysRev.36.383,PhysRev.39.616}. In the quasi-relativistic limit (i.e.\ in the second order in the $v/c$ parameter), the pair interaction of electrons is described by the well-known Breit-Pauli Hamiltonian~\cite{bethe2012quantum}. Along with usual Coulomb term, it also contains the PSOI that features two components:
\begin{equation}
\label{breit}
    H_{\mathrm{Breit}}  =\frac{e \hbar}{4 m^{2} c^{2}} \sum_{i \ne j} \left( \mathbf{E}_{ij} \times \mathbf{p}_{i} + \frac{2 e}{r_{ij}^{3}} \mathbf{r}_{ij} \times \mathbf{p}_{j}\right)\cdot \bm{\sigma}_i \,.
\end{equation}
The first term in Eq.~\eqref{breit} is quite similar to the SOI of Eq.~\eqref{pauliso} in the Pauli-Schrödinger equation, but with the Coulomb field of pair interaction $\mathbf{E}_{ij}$ in place of the external field. Each electron feels the SOI created by the electric field of the other electron. The second term describes the action of the magnetic field, created by  a moving electron, on the spin of the other electron. 

\subsection{Pair SOI in crystals}

The Breit-Pauli Hamiltonian can be applied for electrons in crystals within the $k \cdot p$ method assuming that the electric field $\mathbf{E}_{ij}$ is smooth on the scale of the lattice constant. This results in the replacement of $p$ by a quasi-momentum and the appearance of coefficients that are determined by the band spectrum and basis Bloch states. It is very important that the coefficients arising in the first (electric) and second (magnetic) terms in Eq.~\eqref{breit} are quite different. The ratio of the coefficients depends on the model of the band structure, but our estimates show that the contribution of the magnetic term is negligible.

It is easy to demonstrate this in the $8\times 8$ Kane model~\cite{KANE1957249,voon2009kp}. The electric component of the PSOI equals
\begin{equation}
\label{pse}
    H_{\mathrm{PSOI}} = \frac{\alpha}{\hbar} \sum_{i \ne j} \left(\mathbf{E}_{ij} \times \mathbf{p}_{i} \right)\cdot \bm{\sigma}_{i} \,,
\end{equation}
with
\begin{equation}
	\alpha = \frac{e P^{2}}{3} \left[ \frac{1}{E_{0}^{2}} - \frac{1}{{(E_{0}+\Delta_{0})}^2} \right] \,,
\end{equation}
where $E_{0}$ is the band gap, $P$ is the interband matrix element of the momentum, and $\Delta_{0}$ is the spin-orbit splitting~\cite{winkler}. The magnetic component is
\begin{equation}
	H_\mathrm{B}  = g^{*} \mu_{B} \sum_{i \ne j} \mathbf{B}_{ij} \cdot \bm{\sigma}_{i} \,,
\end{equation}
where $\mu_{B}$ is Bohr's magneton, and the effective $g$-factor is
\begin{equation}
	g^{*} = g_{0} - \frac{2 m_{0}}{\hbar^{2}} \frac{P^{2}}{3}\left[ \frac{1}{E_{0}} - \frac{1}{E_{0} + \Delta_{0}} \right]\,.
\end{equation}
Therefore
\begin{equation}
	\left| \frac{H_{\mathrm{B}}}{H_{\mathrm{PSOI}}} \right| \sim \frac{2 E_{0}}{m^{*} c^{2}}\,,
\end{equation}
where $m^{*}$ is the electron effective mass. It is clear that the ratio of the magnetic PSOI contribution to the electric contribution is extremely small as long as $E_{0} \ll m^{*} c^{2}$. Qualitatively, the result holds under any realistic assumption about the spectrum.

Hence the PSOI Hamiltonian in the crystal has the form of Eq.~\eqref{pse}, with $\alpha$ taken as a material-dependent constant of the same order of magnitude as the constant of the Rashba SOI in Eq.~\eqref{rham}.  The PSOI can be interpreted as the Rashba SOI produced by the Coulomb field $E$ of the interacting electrons. The fundamental difference between them is that the electric field $E$ that defines the Rashba SOI constant $\alpha_{R} = \alpha E$ depends on the electron positions and thus should be determined self-consistently from the full quantum-mechanical solution of the problem.

This model is supported by the calculations of the spin-orbit splitting of the surface electron states formed by the image potential in metals~\cite{0953-8984-16-39-017}. The calculations, based on relativistic multiple-scattering equations, show that the image-potential-induced SOI is correctly described by Eq.~\eqref{pse}. The calculated spectra agree well with recent experiments on the surface of Au(001)~\cite{PhysRevB.94.115412} and on the graphene/Ir(111) interface~\cite{PhysRevLett.115.046801}. The effects of the SOI, induced by the Coulomb fields of electrons in parallel 2D layers, were explored in Ref.~\cite{PhysRevB.84.033305}.

\subsection{When PSOI comes into play}

The PSOI proves important if its amplitude $\alpha$ is large enough. Let us estimate at what $\alpha$ the PSOI significantly modifies the pair interaction by comparing the PSOI energy $E_{\mathrm{PSOI}} \sim \alpha k \frac{e}{r^{2}}$ with the Coulomb part of the e-e interaction, $V \sim \frac{e^{2}}{r}$. The spin-orbit component of the pair interaction dominates when  $\alpha k > e r $, or, estimating the wave-vector $k$ as $1/r$, when $\alpha > e r^{2}$. 

The minimal value of $r$ in this estimate is limited by a characteristic scale $d$ of the model. The scale can be set either by the system geometry, e.g.\ by the thickness of the layer that is considered as a 2D system, or by some physical factors like the Zitterbewegung, which reflects the existence of the second band in the energy spectrum. In any case the scale $d$ is at least of the order of $\SI{10}{\angstrom}$, so we can estimate the minimal value of $\alpha$ as of $\SI{100}{e \angstrom \squared}$.

Take, for instance, GaAs with $\alpha \approx \SI{1}{e \angstrom \squared}$, where this condition fails. However, the required values of $\alpha$ are attainable in such materials as $\mathrm{Bi}_2 \mathrm{Se}_3$~\cite{PhysRevLett.107.096802}, where $\alpha \approx \SI{e3}{e \angstrom \squared}$,  $\mathrm{BiTeI}$~\cite{ishizaka2011giant}, the $\mathrm{BiSb}$ monolayers~\cite{PhysRevB.95.165444}, 2D transition metal dichalcogenides~\cite{manzeli20172d}, graphene with adsorbed heavy elements~\cite{otrokov2018evidence,PhysRevB.99.085411}, perovskites~\cite{PhysRevLett.117.126401} and oxides~\cite{varignon2018new}.

\section{Electrostatics and PSOI in low-dimensional systems}
\label{elstat}

The form of the PSOI Hamiltonian in a specific low-dimensional structure is determined by the configuration of the Coulomb fields therein. In a 1D quantum wire the only source of the PSOI is the Coulomb field of e-e interaction directed normally to the wire, which appears if the axial symmetry of the system is broken~\cite{PhysRevB.95.045138,2017arXiv170700316G,doi:10.1002/pssr.201700256,2018arXiv180410826G}. Such asymmetry can be created, for example, by a proximate metallic gate. In this case, the PSOI results from the image charges induced by 1D electrons on the gate. In 2D systems symmetric at the in-plane reflection the PSOI is created by the in-plane electric field~\cite{PhysRevB.98.115137,10.1016/j.physe.2018.12.028}, whereas in the gated 2D systems the PSOI is created jointly by the in-plane and normal electric fields, the interplay of which leads to rather unexpected results~\cite{2019arXiv190409510G}. Below we discuss the electric field configurations in these systems in more detail to derive the specific form of the PSOI Hamiltonian in each case.

\begin{figure}
    \centering
    \begin{minipage}{0.45\textwidth}
        \centering
        \includegraphics[width=0.9\textwidth]{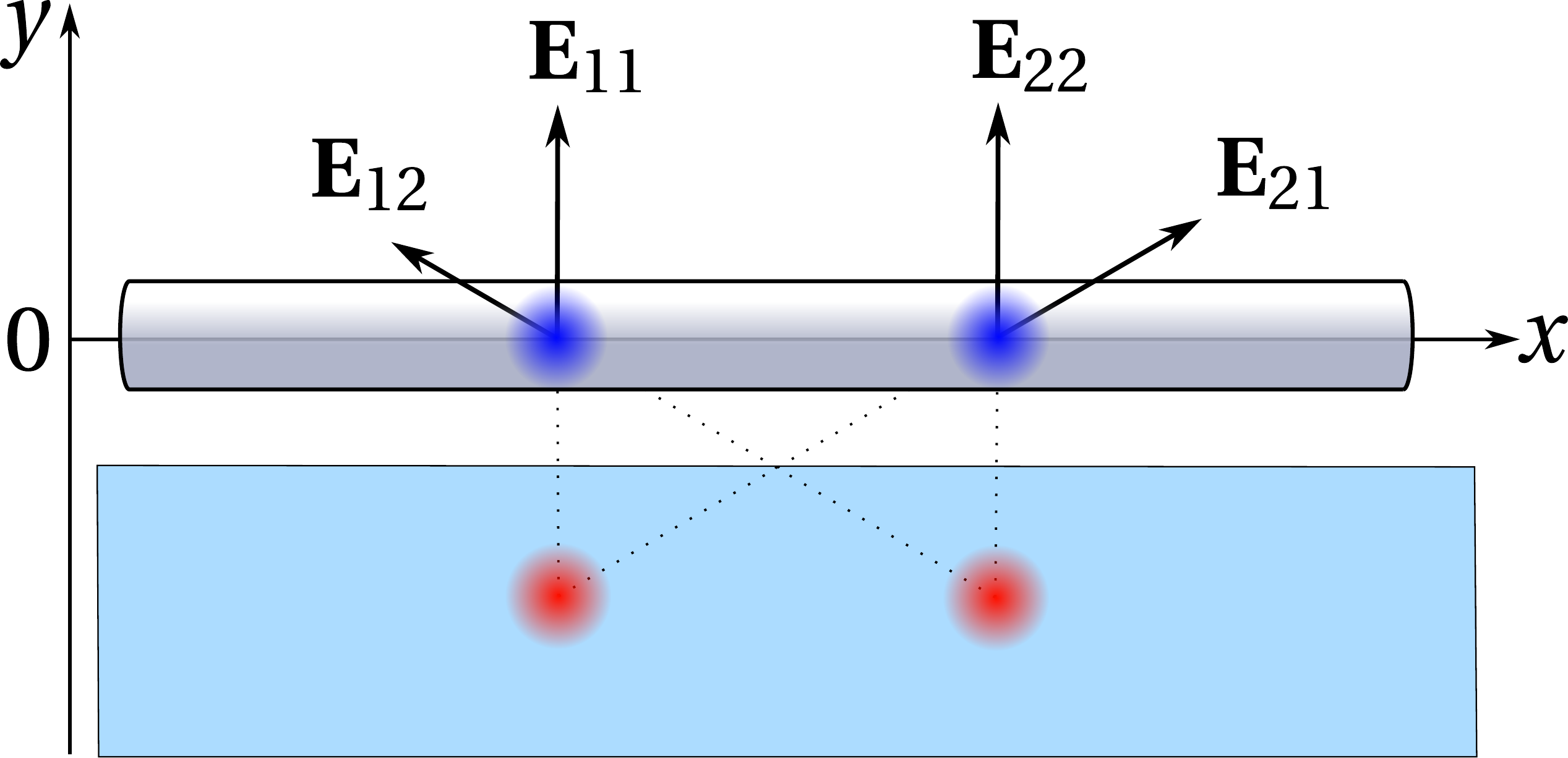}
        \caption{A quantum wire with two electrons that induce mirror charges on the gate. The electric fields acting on each electron from the mirror charges are shown by arrows, whereas the field created directly by a neighboring electron is not shown as it does not contribute to the PSOI in 1D system.\label{fig_3_1}}
    \end{minipage}\hfill
    \begin{minipage}{0.45\textwidth}
        \centering
        \includegraphics[width=0.9\textwidth]{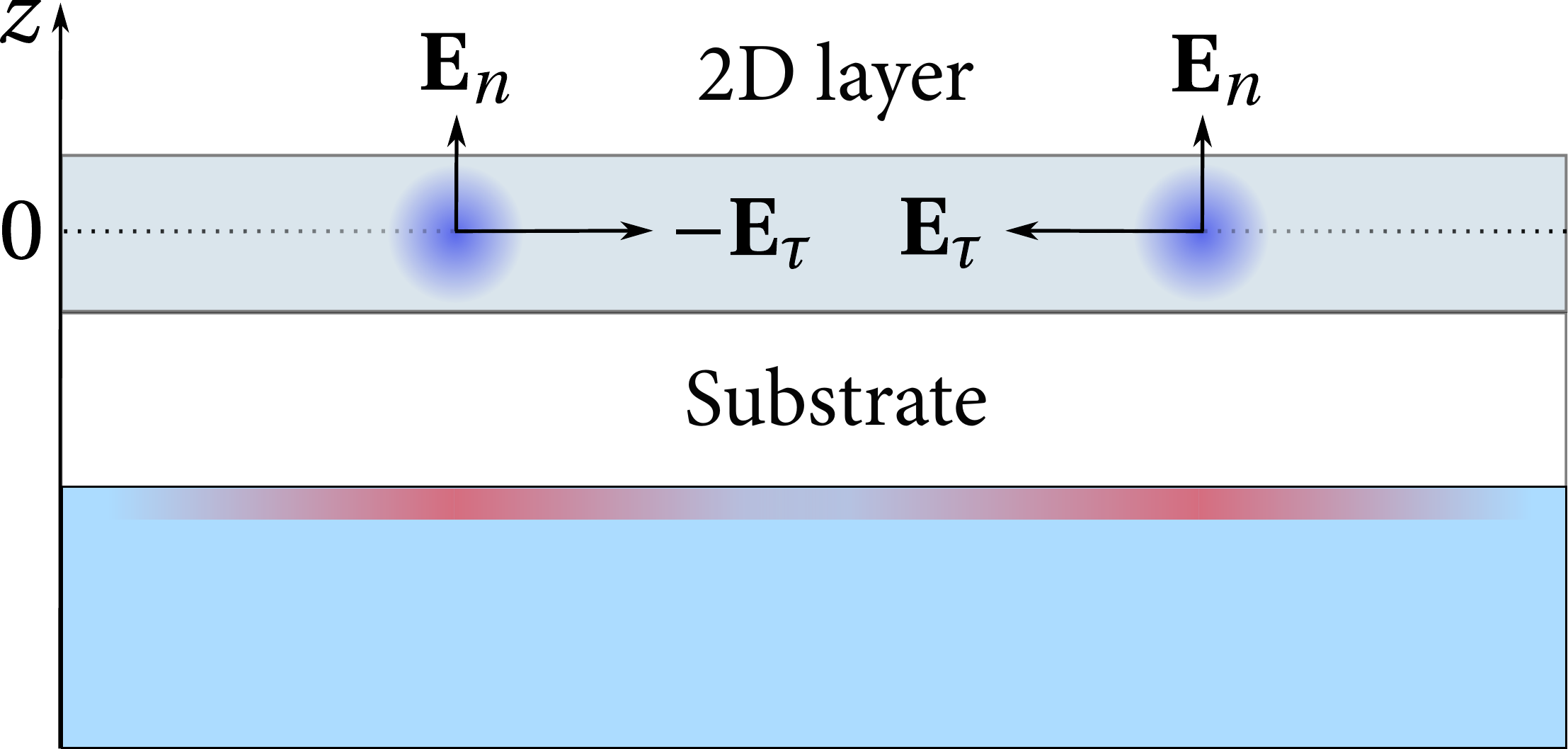} 
        \caption{A 2D layer separated by a weak dielectric substrate from the gate. The electric field acting on each electron consists of the normal ($\mathbf{E}_{n}$) and in-plane ($\mathbf{E}_{\tau}$) component created by a neighboring electron, the polarization charges, and the total charge of the gate.\label{fig_3_2}}
	\end{minipage}
\end{figure}

\subsection{1D quantum wires}
Consider a 1D quantum wire placed parallel to the conductive gate as shown in Fig.~\ref{fig_3_1}. The Coulomb repulsion between two electrons, screened by the image charges, is defined by the e-e interaction potential
\begin{equation}
\label{ee1d}
        U(r) = \frac{e^2}{\epsilon \sqrt{r^2 + d^2}} - \frac{e^2}{\epsilon \sqrt{r^2 + a^2}}\,,
\end{equation}
where $r = x_1 - x_2$ is the relative position of two electrons, $\epsilon$ is a dielectric constant of the bulk material, $d$ is the wire diameter, and $a/2$ is the separation between the wire and the gate. 

The electric field acting on each electron is created by a neighboring electron and the image charges induced on the gate. The specific feature of the 1D case is that the electric field of the neighboring electron, being parallel to the electron momentum, does not create the SOI\@. The only contribution to the PSOI of Eq.~\eqref{pse} comes from the field by the mirror charge of the neighboring electron that has a component normal to the wire,
\begin{equation}
\label{fld1d}
	E_{ij}^{y} \equiv E_n(r) = \frac{e a}{\epsilon{(r^2 + a^2)}^{\frac{3}{2}}}\,.
\end{equation}

The PSOI Hamiltonian has the form~\cite{PhysRevB.95.045138}
\begin{equation}
\label{isoi}  
    H_\mathrm{PSOI} = \frac{\alpha}{2\hbar} \sum_{s_1s_2}  \int \psi^+_{s_1}(x_1) \psi^+_{s_2}(x_2) {[\hat{\mathcal{S}}_{12}, E_n(x_1-x_2)]}_{+} \psi_{s_2}(x_2) \psi_{s_1}(x_1)\,dx_1 dx_2\,,
\end{equation} 
where $\psi_{s}(x)$ is the electron field operator, $\hat{\mathcal{S}}_{12} = (\hat{p}_{x_1} s_1 + \hat{p}_{x_2} s_2)/2$, and the anti-commutator ${[\hat{A}, \hat{B}]}_{+} = \hat{A}\hat{B} + \hat{B}\hat{A}$ is introduced to maintain the hermiticity of the Hamiltonian when projecting Eq.~\eqref{pse} to the 1D subspace. To be specific, we assume $\alpha >0$ from now on.

The normal field $F = F_0 + F_{g}$ created by electron's own image $F_0 = E_n(0)$ and the charge density of the gate $F_{g} = 4 \pi n_{g}/\epsilon$, respectively, contributes to the one-body Rashba SOI via
\begin{equation}
\label{soi1}
    H_\mathrm{RSOI} = \frac{\alpha}{\hbar}\sum_{s}\int \psi^+_{s}(x) F \,\hat{p}_{x} s \psi_{s}(x)\, dx\,.
\end{equation}

When considering a two-body problem, it is convenient to express the PSOI Hamiltonian in the two-particle basis $\{\lvert \uparrow \uparrow  \rangle, \lvert \uparrow \downarrow  \rangle, \lvert \downarrow \uparrow  \rangle, \lvert \downarrow \downarrow  \rangle \}$ as~\cite{2018arXiv180410826G}
\begin{equation}
\label{psoi1d}
    H_\mathrm{PSOI} = \frac{\alpha}{\hbar} \mathrm{diag} \left\{ E_n(r) P, {[E_n(r),p]}_{+}, -{[E_n(r),p]}_{+}, -E_n(r) P \right\} \,,
\end{equation}
with $p = -i \hbar \partial_{r}$, the center-of-mass position $R = (x_1 + x_2)/2$ and corresponding momentum $P = -i \hbar \partial_{R}$. A one-body Rashba SOI can be easily taken into consideration by adding the field $F$ to the normal field  $E_n(r)$ in Eq.~\eqref{psoi1d}.

It is convenient to introduce a dimensionless SOI constant $\tilde{\alpha} = \alpha/e a_{B}^{2}$, with the Bohr radius $a_{B} = \epsilon \hbar^2/m e^2$, and the Rydberg constant in the material $Ry = \hbar^2/2 m a_B^2$. In numerical calculations below we will assume $\tilde{\alpha} \approx 1$ and $a_B \approx \SI{100}{\angstrom}$. These values are close to those attainable in materials like $\mathrm{Bi}_2 \mathrm{Se}_3$~\cite{PhysRevLett.107.096802,manchon2015new}.

\subsection{2D gated layers}
In searching for the most promising 2D system to study the PSOI effects, one has to take into account that in materials with a strong Rashba effect $\epsilon$ is typically large~\cite{manchon2015new}, which severely suppresses the e-e interaction effects. In his seminal work Ref.~\cite{keldysh1979coulomb}, Leonid Keldysh pioneered an idea to study a thin 2D layer placed in a weak dielectric environment to reduce the unwanted bulk dielectric screening of the e-e interaction. Such geometry affects the spatial dependence and strength of the e-e interaction potential, which is now commonly referred to as the Rytova-Keldysh potential~\cite{RevModPhys.90.021001}, giving credit to an earlier work of Natalia Rytova~\cite{rytova}. Decades later, the freely suspended 2D layers became the focus of the intensive experimental activity~\cite{ROSSLER2010861,doi:10.1063/1.5019906}. 

This idea can be further developed by considering a sandwich structure where a 2D layer is separated by a weak dielectric spacer from a charged metallic gate as shown in Fig.~\ref{fig_3_2}. Two-dimensional gated layers attract now a great deal of attention, because they represent a system with a highly tunable Rashba SOI~\cite{PhysRevB.95.165444,PhysRevB.99.104505}. Most importantly, at low distances between the electrons, where the electron pairs are formed,  the e-e interaction and hence the in-plane electric field that gives the leading contribution to PSOI are only weakly screened similar to the Rytova-Keldysh potential. The normal field of the image charges additionally contributes to the PSOI, whereas the field of the charged gate allows one to tune the PSOI effects.

The e-e interaction potential is given by~\cite{2019arXiv190409510G}
\begin{equation}
\label{ee2d}
	U(r) = 2 e^2 \int_{0}^{\infty} \frac{J_{0}(k r)\, dk}{1 + 4 \pi \chi k + \coth (k a/2) }\,,
\end{equation}
where  $\mathbf{r} = \mathbf{r}_1 - \mathbf{r}_2$ stands for the relative in-plane position of the electrons, $J_{0}$ is the Bessel function of the first kind~\cite{olver}, $\chi$ is a 2D susceptibility of the layer~\cite{PhysRevB.84.085406,doi:10.1063/1.5052179}, and $a/2$ is the separation from the gate. The in-plane field is $\mathbf{E}_{\tau}(\mathbf{r}) = \frac{1}{e} \nabla_{r} U$, while the normal field equals
\begin{equation}
\label{effield}
    E_{n}(r) = e \int_{0}^{\infty} \frac{J_{0}(kr) k\, dk}{e^{k a} + 2 \pi \chi k (e^{ka}-1)}\,.
\end{equation}

The 2D susceptibility can be estimated as $\chi \approx \epsilon d/4 \pi$, where $\epsilon$ is the in-plane component of the dielectric tensor of the bulk material, and $d$ is the layer thickness~\cite{PhysRevB.88.045318}. Of particular interest is the case of $a \lesssim 4 \pi \chi$, when the image charges induced on the gate are fully involved. At $r \ll a$, we have
\begin{equation}
\label{nas}
	E_{n}(r) = \frac{e}{2 \pi \chi} \frac{1}{\sqrt{r^2 + a^2}} 
\end{equation}
and 
\begin{equation}
\label{astau}
    \mathbf{E}_{\tau}(\mathbf{r}) = - \frac{e}{2 \pi \chi r} \frac{\mathbf{r}}{r}\,.
\end{equation}
Since the magnitude of the in-plane field is not cut-off by the distance to the gate, its contribution to the PSOI is generally speaking larger than that of the normal field.

The PSOI Hamiltonian in the two-particle basis is given by~\cite{2019arXiv190409510G}
\begin{equation}
\label{PSOI}
    H_{\mathrm{PSOI}} = \frac{\alpha}{2 \hbar}
        \left(
            \begin{matrix}
                \frac{4 E_{\tau}(r)}{r} {(\mathbf{r} \times \mathbf{p})}_{z} && -\xi_{+} + \Xi_{+} && \xi_{+} + \Xi_{+} && 0 \\
                -\xi_{-} + \Xi_{-} && \frac{2 E_{\tau}(r)}{r} {(\mathbf{r} \times \mathbf{P})}_{z} && 0 && \xi_{+} + \Xi_{+} \\
                \xi_{-} + \Xi_{-} && 0 && -\frac{2 E_{\tau}(r)}{r} {(\mathbf{r} \times \mathbf{P})}_{z} && -\xi_{+} + \Xi_{+} \\
                0 && \xi_{-} + \Xi_{-} && -\xi_{-} + \Xi_{-} && -\frac{4 E_{\tau}(r)}{r} {(\mathbf{r} \times \mathbf{p})}_{z}
            \end{matrix}
        \right)\,.
\end{equation}
Here we introduced the center-of-mass position $\mathbf{R} = (\mathbf{r}_{1} + \mathbf{r}_{2})/2$, the momenta $\mathbf{p} =  -i \hbar\nabla_{\mathbf{r}}$ and $\mathbf{P} =  -i \hbar\nabla_{\mathbf{R}}$. Then, $\xi_{\pm} = {[ E_{n}(r),\gamma_{\pm}]}_{+}$, $\Xi_{\pm} = E_{n}(r) \Gamma_{\pm}$, where $\Gamma_{\pm} = P_{y} \pm i P_{x}$ and
\begin{equation}
		\gamma_{\pm} = p_y \pm i p_{x} = \hbar e^{\mp i \phi} \left(\pm \partial_{r} - \frac{i}{r}\partial_{\phi} \right)\,.
\end{equation}
A one-body Rashba SOI can be included in Eq.~\eqref{PSOI} similarly to the 1D case by adding $F$ to the normal field  $E_n(r)$.

\section{Spin-orbit mechanism of electron pairing}
\label{pairing}
Electron pairing is commonly related to the attractive interaction mediated by the crystal lattice or by the many-particle excitations of the electron system~\cite{combescot2015excitons,kagan2013modern}. PSOI leads to a  pure electronic mechanism of the pairing. What is surprising about it, the electron pairing results from the mere motion of electrons in certain spin configurations provided that the PSOI magnitude is high enough. In this section, we discuss the two-electron bound states formed in 1D quantum wires and 2D layers, the pairing conditions, the energy spectrum, the charge and spin density distribution, and give estimates of the binding energy, which can be as large as several milli-electron volts.

\subsection{Electron pairing in quantum wires}

The two-electron wave-function $\Psi(x_1,x_2)$ is a Pauli spinor of the 4-th rank. Owing to the translational invariance, $\Psi(x_1,x_2) = e^{i K R} \psi(r)$, with $\hbar K$ being the total momentum of the pair. The spinor $\psi(r) = {(\psi_{\uparrow \uparrow},\psi_{\uparrow \downarrow},\psi_{\downarrow \uparrow},\psi_{\downarrow \downarrow})}^{\intercal}$ describes only the relative motion of electrons, although its components can in principle depend on $K$ since the binding potential depends on $K$.

The equation of motion for $\psi(r)$ is defined by the total Hamiltonian $H = H_\mathrm{PSOI} + U + T$ that includes the PSOI Hamiltonian of Eq.~\eqref{psoi1d} plus the Coulomb repulsion of Eq.~\eqref{ee1d} and the kinetic energy $T$. For simplicity, we consider a minimal model with quadratic band dispersion. The Schrödinger equation for $\Psi$ leads to the following equations for the spinor components $\psi_{\uparrow \downarrow}$ and $\psi_{\uparrow \uparrow}$~\cite{2018arXiv180410826G}
\begin{equation}
\label{rel}
		\left[ -\frac{\hbar^2}{m} \partial_{r}^2 - 2 i \alpha (F + E_n (r))\partial_{r} -  i\alpha E_n '(r) + U(r) \right] \psi_{\uparrow \downarrow} = \left(\varepsilon_{\uparrow \downarrow} - \frac{\hbar^2 K^2}{4 m}\right)\psi_{\uparrow \downarrow}
\end{equation}
and
\begin{equation}
\label{conv}
		\left[ - \frac{\hbar^2}{m} \partial_{r}^2 + \alpha K E_n (r) + U(r)\right]\psi_{\uparrow \uparrow}
		= \left(\varepsilon_{\uparrow \uparrow} - \frac{\hbar^2 K^2}{4 m} - \alpha K F \right) \psi_{\uparrow \uparrow}\,.
\end{equation}

The equations for the remaining spinor components $\psi_{\downarrow \uparrow}$ and $\psi_{\downarrow \downarrow}$ are obtained from Eqs.~\eqref{rel} and~\eqref{conv}, respectively, by changing the sign before $\alpha$. The equations for the spinor components are uncoupled because the total Hamiltonian is diagonal in the chosen two-particle basis. However, since $\psi(r)$ should be antisymmetric with respect to the particle permutation, both components $\psi_{\uparrow \downarrow}$ and $\psi_{\downarrow \uparrow}$ are mixed in the full solution of the system. The solutions of Eqs.~\eqref{rel}-\eqref{conv} that belong to the discrete spectrum describe the two-electron bound states of principally different nature.

\subsubsection{Relative bound states}
Thus, Eq.~\eqref{rel} describes the relative bound states which arise due to the attractive potential formed by the relative motion of electrons with opposite spins, with the motion of the center-of-mass fully decoupled. 

\begin{figure}
    \centering
    \begin{minipage}{0.45\textwidth}
        \centering
        \includegraphics[width=0.9\linewidth]{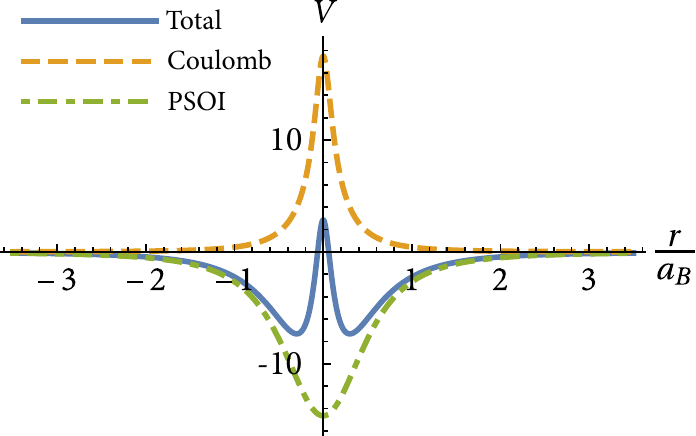}
		\caption{The binding potential for the relative motion $V(r)$ (in $Ry$ units) and its components due to the Coulomb e-e interaction and PSOI\@. The system parameters are $a = 0.8 a_B$ and $d = 0.1 a_B$.\label{fig_4_1}}
    \end{minipage}\hfill
    \begin{minipage}{0.45\textwidth}
        \centering
        \includegraphics[width=0.9\linewidth]{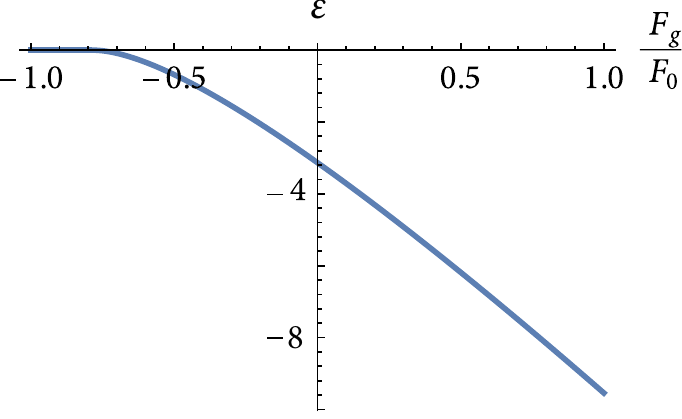}
		\caption{The dependence of the binding energy (in $Ry$ units) on the gate field as calculated numerically from Eq.~\eqref{equ}. Same parameters as in Fig.~\ref{fig_4_1}.\label{fig_4_2}}
	\end{minipage}
\end{figure}

The explicit form of the binding potential is obtained by performing a gauge transformation $\psi_{\uparrow \downarrow}(r) = u(r) e^{-i \phi(r)}$ with
\begin{equation}
\label{gauge}
	\phi(r)= \frac{m \alpha}{\hbar^2} \int_0^{r} \left(F + E_n(\eta)\right)\,d \eta
\end{equation}
that leads to the following equation of motion for the transformed function
\begin{equation}
    \label{equ}
        - \frac{\hbar^2}{m} u'' + V(r) u = \varepsilon u\,,
\end{equation}
where $\varepsilon = \varepsilon_{\uparrow \downarrow} - \frac{\hbar^2 K^2}{4 m} + \frac{m \alpha^2}{\hbar^2} F^2$ is the binding energy. The potential profile
\begin{equation}
    V(r) = U(r) - \frac{m \alpha^2}{\hbar^2}\left[E_n^2(r) + 2 F E_n(r) \right]
\end{equation}
is illustrated in Fig.~\ref{fig_4_1}, with separately shown contributions from the Coulomb interaction and PSOI\@.
If a sufficiently large PSOI prevails over the Coulomb repulsion so that the binding potential $V(r)$ becomes globally attractive, i.e.  $\int V(r)\, dr <0$, then a bound state appears in the spectrum~\cite{simon1976bound}. 
The binding energy of the relative state is estimated as~\cite{2018arXiv180410826G} 
\begin{equation}
	\label{bind}
	|\varepsilon| = 2 Ry {\left[ \frac{2 \tilde{\alpha}^2}{{(a/a_B)}^3} \left(\frac{F}{F_0} + \frac{3 \pi}{32}\right) - \log \frac{a}{d} \right]}^2\,.
\end{equation}
The bound state appears in the spectrum as soon as the expression in the square brackets is positive. This can always be achieved by increasing $F$, that is by applying voltage to the gate. Increasing the voltage facilitates the pairing by increasing the binding energy, which is illustrated by Fig.~\ref{fig_4_2}. Equation~\eqref{bind} leads to the binding energies in the meV range.

The wave function of the relative bound states in quantum wires
\begin{equation}
	\psi(r) = {\left( 0, e^{-i \phi(r)}, -e^{i \phi(r)}, 0 \right)}^{\intercal} u(r)
\end{equation}
is of the mixed singlet-triplet type.

\subsubsection{Convective bound states}
Contrary to the relative bound states, the attraction in Eq.~\eqref{conv} arises exactly because of the center-of-mass motion. This leads to the convective bound states. In 1D systems, both types of the bound states always lie in the gap below the conduction band bottom.
 
The attractive potential $V(r) = \alpha K E_n(r)$  is proportional to the total momentum of the pair $K$, the sign and magnitude of which determine the spin structure and spectrum  of the bound states. Convective states arise for $\tilde{\alpha} |K| > (1 + a - d)$, with all variables normalized to Bohr’s radius. Large negative $K$ supports the state of $\psi_{\uparrow \uparrow}$, and positive $K$ supports $\psi_{\downarrow \downarrow}$~\cite{2018arXiv180410826G}. Thus, in 1D quantum wires the convective bound states are formed by electrons with parallel spin orientation locked to the direction of the center-of-mass momentum. Note that in this case the gate field $F$ affects neither the binding potential nor the binding energy $\varepsilon$, which is nonetheless tunable by the magnitude of $K$, that is by the current in the wire.

\subsection{Electron pairing in 2D systems}

PSOI leads to electron pairing in 2D electron systems similarly to the 1D case~\cite{PhysRevB.98.115137,10.1016/j.physe.2018.12.028,2019arXiv190409510G}. However, the spin structure and spectrum of the bound states in 2D layers are quite different from those found in 1D quantum wires because of the different configuration of the electric field that defines the PSOI Hamiltonian.

Despite these differences, the bound states can still be classified according to the nature of the motion of electrons that gives rise to the PSOI, leading again to the picture of the relative and convective states.

\subsubsection{Convective bound states}
The convective states were investigated in detail for symmetric (non-gated) 2D systems in the case of a purely Coulomb in-plane field~\cite{PhysRevB.98.115137}, where an exact analytic solution is possible, and for realistic screening in the layer of material~\cite{10.1016/j.physe.2018.12.028}. In both cases, the convective states are formed by electrons with opposite spins. The equations for the corresponding spinor components of the two-electron wave-function $\Psi(\mathbf{r}_1,\mathbf{r}_2) = {\left(\Psi_{\uparrow \uparrow},\Psi_{\uparrow \downarrow},\Psi_{\downarrow \uparrow},\Psi_{\downarrow \downarrow}\right)}^{\intercal} = e^{i \mathbf{K} \cdot \mathbf{R}} \psi(\mathbf{r},\mathbf{K})$ follow from the PSOI Hamiltonian of Eq.~\eqref{PSOI} with $E_n \equiv 0$ and read as 
\begin{equation}
\label{conv_2d}
	\left[- \frac{\hbar^2}{m} \nabla^2_{\mathbf{r}} - \frac{\hbar^2}{4m} \nabla^2_{\mathbf{R}} + U(\mathbf{r}) + \frac{\alpha}{\hbar} \frac{E_{\tau}(\mathbf{r})}{r} {(\mathbf{r} \times \mathbf {P})}_z \right] \Psi_{\uparrow \downarrow} 
	= \varepsilon_{\uparrow \downarrow} \Psi_{\uparrow \downarrow}\,,
\end{equation}
and similarly for $\Psi_{\downarrow \uparrow}$ with a sign change before $\alpha$.

\begin{figure}
    \centering
    \begin{minipage}{0.45\textwidth}
        \centering
        \includegraphics[width=0.9\textwidth]{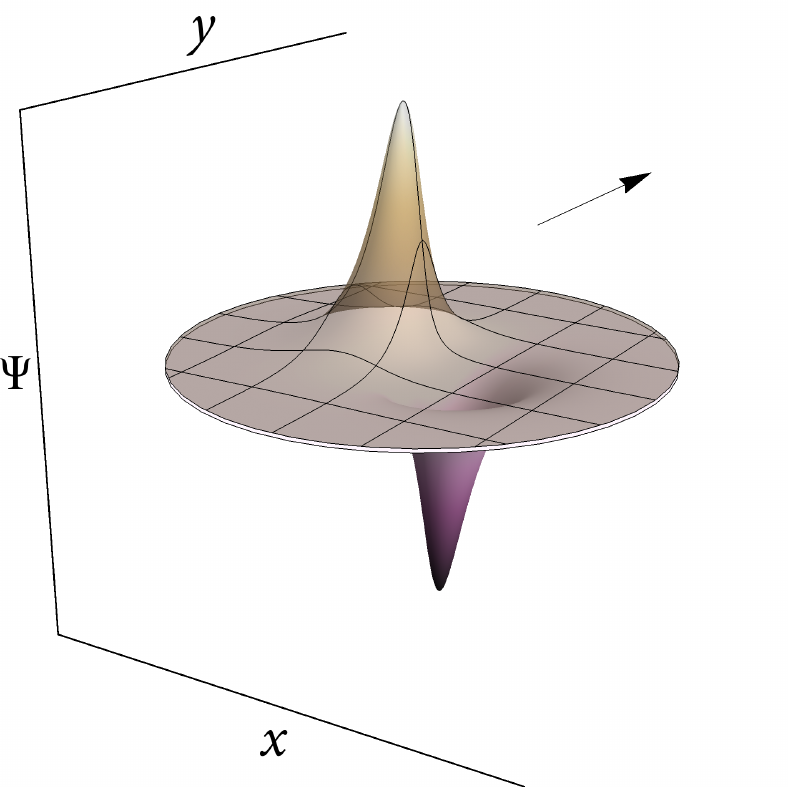} 
        \caption{Two spinor components of the convective state wave function of Eq.~\eqref{convspinor} (shown in different color) as functions of relative coordinates. The arrow shows the direction of vector $\mathbf{K}$. \label{fig_4_3}}
	\end{minipage}\hfill
    \begin{minipage}{0.45\textwidth}
        \centering
        \includegraphics[width=0.9\textwidth]{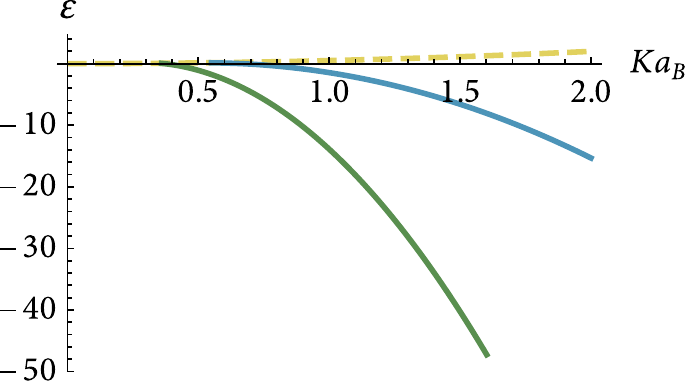}
        \caption{The energy levels of the ground convective state and first excited two-electron state (solid lines) as well as the kinetic energy of the center of mass (dashed line) in $Ry$ units vs $K a_B$, calculated for a symmetric 2D system with the Rytova-Keldysh screening~\cite{10.1016/j.physe.2018.12.028}. \label{fig_4_4}}
    \end{minipage}
\end{figure}

The last term on the left hand side of the equation is the binding potential created by PSOI\@. The potential, proportional to the center-of-mass momentum $\mathbf{K}$, is strongly anisotropic, because the rotational symmetry in the plane is broken by the presence of a preferred direction along $\mathbf{K}$. Correspondingly, the wave function features a non-trivial angular dependence, which has the following form for a purely Coulomb in-plane field,
\begin{equation}
    \label{convspinor}
        \Psi(\mathbf{r},\mathbf{R}) = {\left( 0, ce_0 \left(\frac{\phi}{2},2 \tilde{\alpha} K a_B \right) g(r), -ce_0 \left(\frac{\phi + \pi}{2},2 \tilde{\alpha} K a_B \right) g(r), 0\right)}^{\intercal} e^{i \mathbf{K} \cdot \mathbf{R}}\,,
\end{equation}
with $\phi$ being the polar angle measured from the $\mathbf{K}$ direction, the Mathieu function $ce_{0}(z,q)$~\cite{olver}, and the radial part $g(r)$ given in Ref.~\cite{PhysRevB.98.115137}. Both components $\psi_{\uparrow \downarrow}(\mathbf{r},\mathbf{K})$ and $\psi_{\downarrow \uparrow}(\mathbf{r},\mathbf{K})$ are shown in Fig.~\ref{fig_4_3}. Equation~\eqref{convspinor} represents a mixed singlet-triplet state.

The convective states appear in the spectrum for the center-of-mass momentum $K$ exceeding a critical value. As illustrated by Fig.~\ref{fig_4_4}, the binding energy increases with $K$ so quickly that the total energy of the pair starts to decrease with $K$, even leading to the negative effective mass of the electron pair in some interval of $K$. Note that thanks to a weak dielectric screening in the layer, the binding energy increases by a factor of about $\epsilon$ as compared to the case of a pure Coulomb field.

\subsubsection{Relative bound states}

In symmetric 2D systems, the relative bound states represent a degenerate pair of triplet-like states formed by electrons with parallel spins locked to the angular momentum direction. The equations for the corresponding spinor components are
\begin{equation}
\label{rel_2d}
	\left [- \frac{\hbar^2}{m} \nabla^2_{\mathbf{r}} - \frac{\hbar^2}{4m} \nabla^2_{\mathbf{R}} + U(\mathbf{r}) + \frac{2\alpha}{\hbar} \frac{E_{\tau}(\mathbf{r})}{r} {(\mathbf{r} \times \mathbf {p})}_z \right ] \Psi_{\uparrow \uparrow}
	= \varepsilon_{\uparrow \uparrow} \Psi_{\uparrow \uparrow}
\end{equation}
and similarly for $\Psi_{\downarrow \downarrow}$, but with $\alpha \to -\alpha$. It is seen that in symmetric 2D systems the center-of-mass motion is fully decoupled from the relative motion of electrons and hence has no impact on the structure and spectrum of the relative states.

The lowest-lying states correspond to the minimum angular momentum $l = \pm 1$. They are
\begin{equation}
\label{deg1}
	\Psi(\mathbf{r}) = {\left(u(r) e^{- i \phi},0,0,0\right)}^{\intercal}
\end{equation}
and 
\begin{equation}
\label{deg2}
	\Psi(\mathbf{r}) = {\left(0,0,0,u(r) e^{i \phi}\right)}^{\intercal}
\end{equation}
with the radial wave-function $u(r)$ determined from the Schr\"odinger  equation
\begin{equation}
\label{rad}
		\left[ -\frac{\hbar^2}{m} \left( \frac{d^{2}}{d r^{2}} + \frac{1}{r} \frac{d}{d r} - \frac{1}{r^2} \right) +V(r) - 2 \alpha \frac{E_{\tau}(r)}{r} \right] u(r) = \varepsilon u(r) \,.
\end{equation}

The last term on the left hand side of Eq.~\eqref{rad} is exactly the binding potential produced by PSOI\@. Since the short-range asymptotic behavior of $E_{\tau}(r)$ is given by Eq.~\eqref{astau}, the bound states are formed by the attractive potential of $\propto -\frac{\alpha}{\chi r^{2}}$.

Here we face a fundamental problem inherent in a single-band treatment of PSOI effects. The thing is that the attractive $-1/r^{2}$ potential is a singular potential~\cite{RevModPhys.43.36}, notoriously known for several decades largely because of its widespread occurrence in quantum physics. The $-1/r^{2}$ potential is encountered in the three-body problem in nuclear physics~\cite{Efimov:1971zz}, in the context of the point-dipole interactions in molecular physics~\cite{PhysRev.153.1}, and when studying the attraction of atoms to a charged wire~\cite{PhysRevLett.81.737}. At the same time, it leads to some pathological properties of the solutions of the Schr\"odinger equation. Thus, a discrete orthogonal set of eigenfunctions with its eigenvalues is not defined by a requirement that the solutions be square integrable, leaving one with a continuum set of bound states with arbitrary energy $\varepsilon <0$. Attempts to fix this by requiring additionally that eigenfunctions be orthogonal fail, as the spectrum of the bound states nonetheless remains unbounded below, so there is no ground state~\cite{PhysRev.80.797}. This is a particular case of the long-standing problem of falling to the center~\cite{landau1958course}, which recently was revived for graphene~\cite{Wang734}. The root of the problem is that while the Hamiltonian with the $-1/r^{2}$ potential is symmetric, it is not self-adjoint~\cite{Meetz1964}. A large number of regularization techniques was developed to treat the problem~\cite{PhysRevLett.85.1590,PhysRevA.64.042103,PhysRevA.76.032112}. In essence, they are based on introducing a short-distance cut-off~\cite{PhysRevD.48.5940}, which is considered as a phenomenological parameter. 

A regularization of the binding potential can be provided by mechanisms such as the Zitterbewegung of electrons in crystalline solids or by a natural cutting-off due to averaging the three-dimensional quantities across the layer thickness. By imposing the cut-off at the layer thickness, the binding energy is estimated as~\cite{10.1016/j.physe.2018.12.028}
\begin{equation}
\label{spec0}
	|\varepsilon| = 2 Ry \frac{x_{1}^{2} (\lambda)}{{(d/a_{B})}^{2}} \,,
\end{equation}
with $x_{1} (\lambda)$ being the first zero of the Macdonald function $K_{i \lambda}(x)$, and the attraction amplitude defined as
\begin{equation}
	\lambda = \sqrt{\frac{4 \tilde{\alpha}}{d/a_{B}} - 1} \,.
\end{equation}
Eq.~\eqref{spec0} leads to the magnitude of $|\varepsilon|$ of several milli-electron volts.

\begin{figure}
    \centering    
    \begin{minipage}{0.45\textwidth}
        \centering
        \includegraphics[width=1\textwidth]{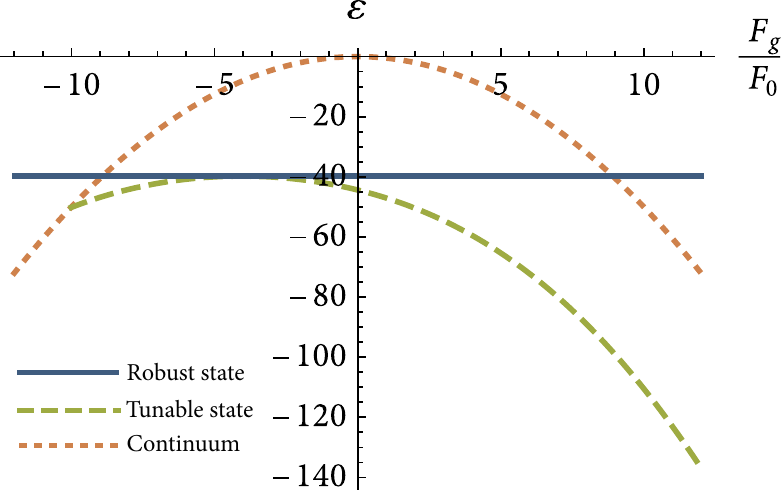} 
        \caption{The binding energy (in $Ry$ units) of the robust and tunable two-electron bound states vs.\ the normalized gate field $F_g/F_0$. Additionally, the position of the conduction band  bottom is shown, from which the binding energy is measured.\label{fig_4_5}}
	\end{minipage}\hfill
    \begin{minipage}{0.45\textwidth}
        \centering
        \includegraphics[width=0.9\textwidth]{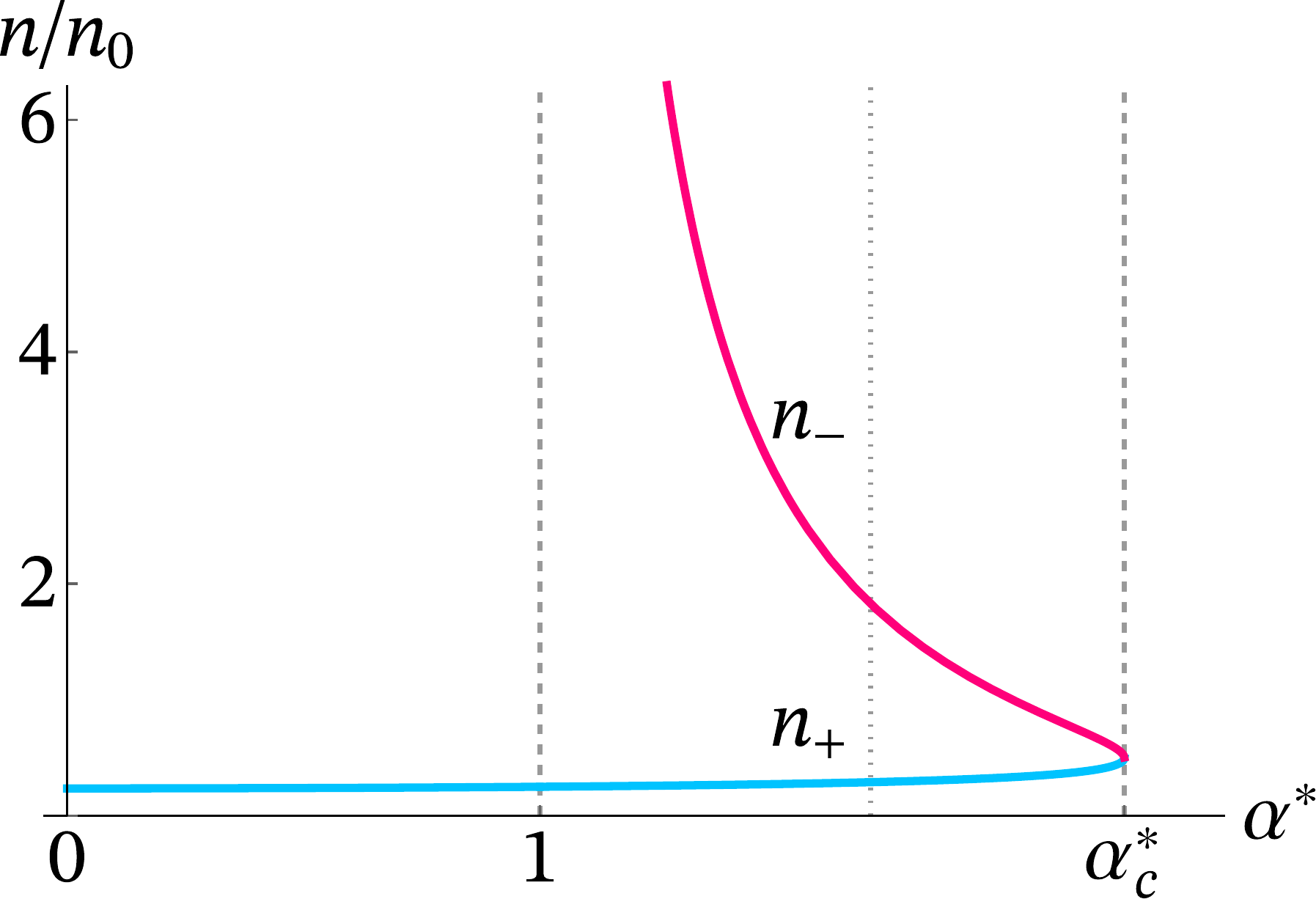}
        \caption{Mean electron density as a function of the PSOI amplitude. Two branches of the solution of Eq.~\eqref{branches} are shown in different color. The Coulomb repulsion amplitude is such that $\mathfrak{v}/\pi \hbar v_F=1$.
		\label{fig_5_1}}
    \end{minipage}
\end{figure}

The normal field $E_n(r)$ that appears in the presence of the gate lifts the twofold spin degeneracy of the relative bound state. The leading role in forming the two-electron bound states still belongs to the in-plane component $E_{\tau}(r)$. However, the interplay of $E_{\tau}(r)$ and $E_n(r)$ results in a dramatic rearrangement of the bound states. The problem is complicated since in the presence of both field components the relative motion is no longer decoupled from the motion of the center of mass. The formation and structure of relative states were investigated for the electron pairs with zero total momentum~\cite{2019arXiv190409510G}. The ground state of an electron pair at rest splits into two states of quite different properties. A tunable state
\begin{equation}
    \label{BS}
        \Psi(\mathbf{r}) = {\left(v(r)e^{- i \phi}, w(r), -w(r), v(r) e^{i \phi}\right)}^{\intercal}
\end{equation}
has a larger binding energy that grows with a gate voltage, with its orbital and spin structure changing continuously. This state disappears as soon as its level crosses the conduction band bottom, which happens at large negative voltage applied to the gate. Surprisingly, there appears also a robust state
\begin{equation}
    \label{prot}
            \Psi(\mathbf{r}) = {\left(u(r)e^{- i \phi},0,0, -u(r) e^{i \phi}\right)}^{\intercal}\,,
\end{equation}
on the orbital and spin structure of which the gate voltage has no effect. Its energy level crosses the bottom of the conduction band at sufficiently high gate voltage of any sign, but the robust state remains bound and localized even in the continuum of band states. This is illustrated by Fig.~\ref{fig_4_5}, where the energy levels of the robust state and tunable bound state are plotted against the normalized field of the gate.

\section{PSOI in a many-electron system}
\label{transport}

Besides the electron pairing, in a many-electron system there appears another strong effect due to the PSOI\@. This is an instability of a homogeneous electron system with respect to the density fluctuations that, too, arises at sufficiently strong PSOI but develops on a large spatial scale~\cite{PhysRevB.95.045138}. It is clear that the behavior of the many-electron system depends on both effects, the relative role of which is yet to be determined. In this section we focus on the PSOI-driven instability of the strongly correlated 1D electron liquids and on the PSOI signatures in electron transport in quantum wires~\cite{2017arXiv170700316G,doi:10.1002/pssr.201700256}.

The instability of the many-electron system is due to a qualitatively new property of PSOI, as compared to the single-electron Rashba SOI\@. The PSOI directly depends on the electron density via the magnitude of the Coulomb electric fields that produces PSOI\@. This dependence creates an efficient mechanism for the density fluctuations to grow, which under certain conditions can result in a radical transformation of the ground state.

The mechanism is as follows. The electron density fluctuation increases the Coulomb electric field that produces the SOI\@. The increased SOI lowers the conduction band bottom and hence the electron energy within the fluctuation region. This leads to the avalanche-like electrons inflow from adjacent regions or reservoirs to the fluctuation region. Thus the density fluctuation once appeared starts to grow.

\subsection{Qualitative treatment of the instability}

Let us determine the electron density in a single-mode quantum wire parallel to a proximate metallic gate, as shown in Fig.~\ref{fig_3_1}, in the case of a fixed chemical potential $\mu$. Let us simplify matters by considering first a mean-field theory with the electron density uniformly distributed along the wire.

Within the mean-field theory, the energy of a single-particle state with wave vector $k$ and spin index $s$ is given by
\begin{equation}
   \varepsilon_{ks} = \frac{\hbar^2}{2m}[{(k + s\,k_\mathrm{so})}^2 - k_\mathrm{so}^2] + \mathfrak{v} \,n\,.
   \label{energy_functional}
\end{equation}
The Coulomb interaction energy is $\mathfrak{v} = 2\frac{e^2}{\epsilon}\ln (a/d)$, and the SOI wave vector is $k_\mathrm{so} = \frac{\alpha m}{\hbar^2} F$. Our Kunststück is in the relation between the normal electric field and the electron density:
\begin{equation}
\label{mean_fld}
    F = 2 n \frac{e}{\epsilon a}\,.
\end{equation}
Equation~\eqref{energy_functional} gives the Fermi wave vectors of $k_F^{(s)} = -s k_{so}\pm {(k_{so}^2 + \frac{2m}{\hbar^2}(\mu - \mathfrak{v}\,n))}^{\frac 12}$. Note that $n = \sum_{s} \int \frac{dk}{2 \pi}$ to obtain the equation for the electron density:
\begin{equation}
   n = \frac{2}{\pi}\sqrt{{\left(\frac{2\alpha me}{\epsilon \hbar^2 a}\right)}^2 n^2 + \frac{2m}{\hbar^2}(\mu - \mathfrak{v} n)}\,.
   \label{instab_n}
\end{equation}

It is convenient to introduce the unperturbed electron density $n_0 = \sqrt{8m\mu}/\pi \hbar$ and Fermi velocity $v_F = \sqrt{2\mu/m}$, as well as the dimensionless SOI amplitude $\alpha^{*} = \frac{4}{\pi}\frac{\tilde{\alpha}}{(a/a_B)}$.
The solution of Eq.~\eqref{instab_n} has two branches,
\begin{equation}
   n_{\pm}(\alpha^{*}) = \frac{n_0}{1 - {\alpha^{*}}^2}\left(-\frac{2\mathfrak{v}}{\pi\hbar v_F} \pm \sqrt{1 - {\alpha^{*}}^2 + {\left(\frac{2\mathfrak{v}}{\pi\hbar v_F}\right)}^2}\right)\,,
   \label{branches}
\end{equation}
which points out to a possible bistability.

\begin{figure}
    \centering
    \includegraphics[width=0.9\textwidth]{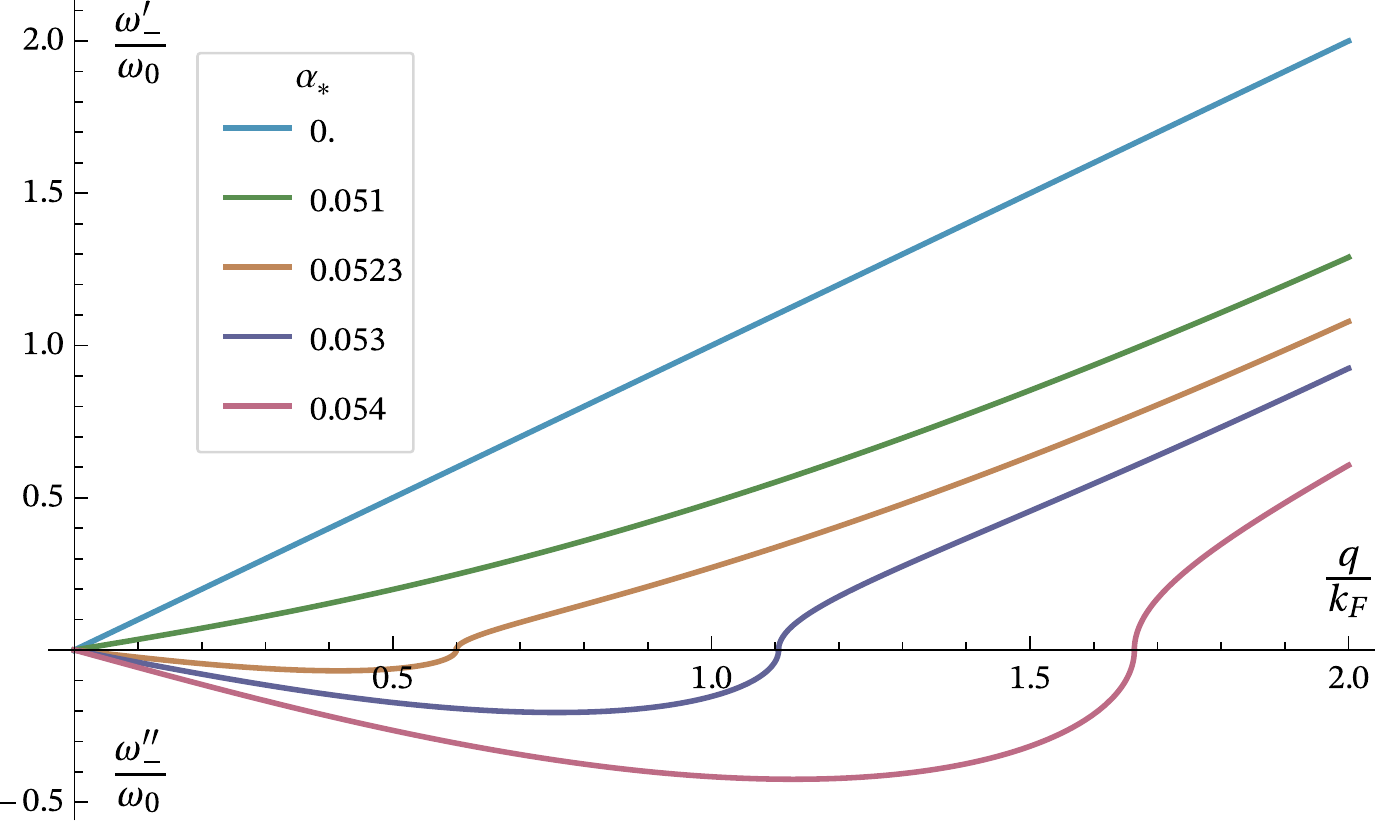} 
    \caption{The real ($\omega'_{-}$) and imaginary ($\omega''_{-}$) parts of the frequency of the collective excitation as a function of the wave vector for several values of the PSOI constant. The frequency is normalized to $\omega_0 = v_F k_F$.\label{fig_5_2}}
\end{figure}

Figure~\ref{fig_5_1} shows an \textit{S}-type dependence of the electron density on the PSOI amplitude. The solution is unique for weak PSOI at $\alpha^{*} < 1$, but there appears a second solution as soon as $1 < \alpha^{*} < \alpha^{*}_c$, where
\begin{equation}
   \alpha^{*}_c = \sqrt{1 + {\left(2\mathfrak{v}/\pi\hbar v_F\right)}^2}\,.
\end{equation}
At $\alpha^{*} > \alpha^{*}_c$, Eq.~\eqref{instab_n} has no solution at all within this simplified model. This behavior of $n(\alpha^*)$ indicates a tendency to form a new ground state at $\alpha^*>\alpha^*_c$, which may be spatially inhomogeneous or may have a strongly correlated structure emerging due to the strong PSOI\@.

The critical value of $\alpha^{*} \approx 1$ corresponding to the onset of the instability with diverging electron density occurs when the distance between the gate and the wire is as small as $a \approx a_B$ provided that the material-dependent SOI constant $\tilde{\alpha}$ is of the order of unity.

\subsection{Collective modes of a 1D electron system with PSOI}
A microscopic model of the 1D electron system with PSOI based on the Hamiltonian of Eq.~\eqref{isoi} was solved by bosonization~\cite{PhysRevB.95.045138,2017arXiv170700316G} and alternatively by solving the equations of motion for the Wigner function in the random phase approximation~\cite{PhysRevB.95.045138,doi:10.1002/pssr.201700256} for the case of the fixed mean electron density $n$.

The spectrum of collective excitations has two branches with frequencies $\omega_{\pm}$ given by
\begin{equation}
\label{dispersion}
    {\left(\frac{\omega_{\pm}}{q v_F}\right)}^2 = 1 + \left( \tilde{U}_q -\alpha_*^2 \tilde{\mathcal{F}} \tilde{E}_q \right) \pm \sqrt{ {\left(\tilde{U}_q - \alpha_*^2 \tilde{\mathcal{F}} \tilde{E}_q \right)}^2 + \alpha_*^2 \tilde{E}_q^2}\,.
\end{equation}
Here we introduced the Fermi velocity $v_F = \hbar k_F/m$, a dimensionless SOI amplitude $\alpha_* = \tilde{\alpha}/(\pi r_s)$ and the interaction parameter $r_s = 1/(2 n a_B)$. The Fourier-transformed Coulomb interaction $U_q = 2(e^2/ \epsilon) \left(K_0(qd) - K_0(qa)\right)$ and normal field $E_q = 2 (e/\epsilon) |q| K_1(|q|a)$ are normalized according to $\tilde{U}_q = U_q/(\pi \hbar v_F)$ and $\tilde{E}_q = \epsilon E_q/(e n_0)$. The mean electric field $\mathcal{F} = F + n E_{q = 0}$ that includes the contribution from the mean electron density is normalized as $\tilde{\mathcal{F}} = \epsilon \mathcal{F}/(e n_0^2)$.

Of most interest is the behavior of the mode $\omega_{-}$ as $\alpha_*$ grows. This is illustrated by Fig.~\ref{fig_5_2} for a quantum wire placed sufficiently close to the gate ($a = 0.4 a_B$, $d = 0.08 a_B$, and $r_s = 0.6$). Increasing PSOI suppresses the mode velocity, which turns to zero for some critical value of PSOI amplitude,
\begin{equation}
    \label{alphat}
        \alpha_c(q) = \dfrac{\sqrt{1 + 2\tilde{U}_q}}{\sqrt{\tilde{E}_q^2 + 2\tilde{\mathcal{F}}\tilde{E}_q}}\,.
\end{equation}
At $\alpha_* > \alpha_c$ the mode frequency even becomes imaginary, i.e.\ the mode acquires a positive increment, which means that the excitations lose their stability. It is important that increasing $\alpha_*$ first leads to the instability in the long-wave region, where the increment has a maximum as a function of $q$. The charge stiffness 
\begin{equation}
    \label{stiffness}
        \varkappa = \pi\hbar v_F (1 + 2\tilde{U}_{0}) \left( 1 - \frac{\alpha_*^2}{\alpha_c^2(0)} \right)
\end{equation}
also turns to zero at $\alpha_* = \alpha_c(0)$, which points to the instability in the charge sector.

\subsection{The spin-charge separation breaking and PSOI signatures in transport}

In the absence of PSOI, $\omega_{+}$ and $\omega_{-}$ branches of excitations of Eq.~\eqref{dispersion} correspond respectively to purely charge excitations (plasmons) and purely spin excitations (spinons).

The spin-charge separation (SCS) between them is a hallmark of the Tomonaga-Luttinger liquid~\cite{giamarchi2003quantum}. In the presence of the single-particle Rashba SOI, the SCS is respected in strictly 1D systems. However, the PSOI violates the SCS in 1D single-mode quantum wires, which leads to the formation of new collective excitations with intertwined spin and charge degrees of freedom~\cite{PhysRevB.95.045138,doi:10.1002/pssr.201700256,2017arXiv170700316G}.

The spin-charge composition of the collective mode is quantitatively characterized by a spin-charge separation parameter $\xi$ that is defined via the relative weight of the electron densities with spin up $n_{\uparrow}$ and down $n_{\downarrow}$ in each mode, and can be directly expressed via the mode phase velocity $v_{\pm} = \omega_{\pm}/q v_F$ as
\begin{equation}
	\xi_{\pm} = \frac{n_{\uparrow} + n_{\downarrow}}{n_{\uparrow} - n_{\downarrow}} = \frac{v_{\pm} - v_{\pm}^{-1}}{\alpha_* \tilde{E}_q} \,.
\end{equation}

\begin{figure}
    \centering
    \begin{minipage}{0.45\textwidth}
        \centering
        \includegraphics[width=1\textwidth]{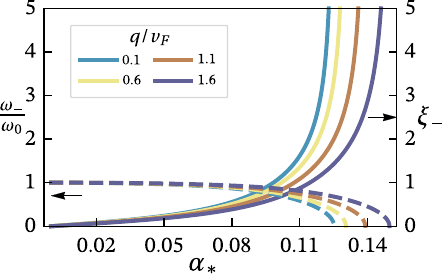}
        \caption{The SCS parameter (solid line) and normalized phase velocity (dashed line) for the $\omega_{-}$  branch of collective excitations as a function of the PSOI amplitude for several $q$.}
		\label{fig_5_3}
    \end{minipage}\hfill
    \begin{minipage}{0.45\textwidth}
        \centering
        \includegraphics[width=1\textwidth]{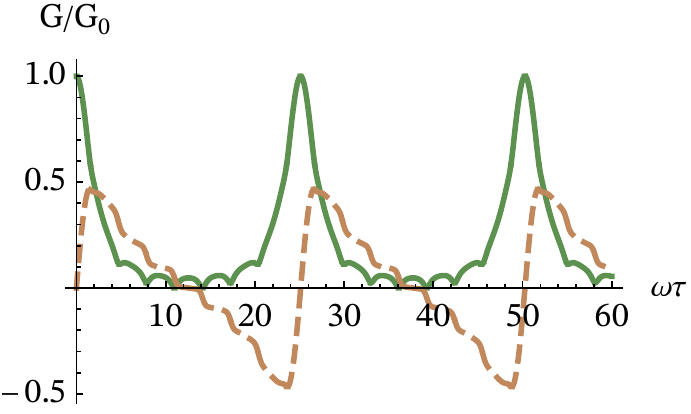} 
        \caption{The real (solid line) and imaginary (dashed line) parts of the admittance versus frequency.}
		\label{fig_5_4}
	\end{minipage}
\end{figure}

At $\alpha_* = 0$ the SCS parameter $\xi_{-} = 0$, which means that the branch $\omega_{-}$ corresponds to a purely spin excitation ($n_{\uparrow} = - n_{\downarrow}$) with the energy dispersion $\omega_{-} = v_F q$ not renormalized by the interactions. As $\alpha_* \to \alpha_c(q)$, the velocity goes to zero $v_{-}(q) \to 0$, whereas the SCS parameter diverges, $\xi_{-} \to \infty$. Consequently, at the instability threshold $\alpha_* = \alpha_c(q)$ the collective mode $\omega_{-}$ turns into a purely charge excitation ($n_{\uparrow} = n_{\downarrow}$). The transformation of the mode spin-charge structure with the change in the PSOI amplitude $\alpha_*$ is shown in Fig.~\ref{fig_5_3}.  

The problem of identification of the SCS breaking in 1D systems with PSOI can be solved by purely electrical measurements of the dynamic conductance (admittance) of the finite 1D system, i.e.\ a 1D quantum wire coupled to leads. This is based on the fact that both spin-charge-mixed modes of a system with broken SCS convey the electric charge and therefore contribute to the electric response of the system.

The analysis of the ac-conductance is known to be a powerful tool for extracting and studying the e-e interaction effects in 1D electronic systems. The real and imaginary parts of the ac-conductance~\cite{Cuniberti_1996,PhysRevB.57.1515,PhysRevLett.81.1925} and especially the ballistic resonances of the admittance~\cite{PhysRevB.58.13847,Sablikov2000} allow one to determine the interaction parameters in the Luttinger liquid, and even to extract the effect of the short-range electron correlations~\cite{PhysRevB.61.12766}. In recent years, this technique has been strongly developed and successfully  applied to the study of electron liquid in carbon nanotubes in the terahertz range~\cite{Zhong2008,4665791,Chudow2016}. However, such a method has not yet been applied to spin effects, possibly with the exception of observing mixed collective modes in the chiral edge channels in quantum-Hall systems with filling factor 2~\cite{Bocquillon2013}.

The ac-transport in the 1D quantum wire coupled to leads was studied in Ref.~\cite{2017arXiv170700316G} to reveal the PSOI signatures in the frequency dependence of the admittance $G_{\omega}$ that allow one to determine the velocities of both branches of collective excitations as well as their spin-charge structure. For a practically important case of a quantum wire placed close to the gate to enhance the PSOI effects, the screened Coulomb e-e interaction becomes short-ranged so that the mode phase velocities $v_{\pm}$ do not depend on the wave-vector. Under these conditions, the admittance is found to be
\begin{equation}
\label{admit}
    \frac{G_{\omega}}{G_0} = \frac{1 - v_{-}^2}{v_{+}^2 - v_{-}^2} \frac{1}{1 - i v_{-} \tan \frac{\omega \tau}{2 v_{-}}} + \frac{v_{+}^2 - 1}{v_{+}^2 - v_{-}^2} \frac{1}{1 - i v_{+} \tan \frac{\omega \tau}{2 v_{+}}}\,.
\end{equation}
Here the conductance quantum is $G_0 = 2e^2/h$, the characteristic transit time is $\tau = L/v_F$, with $L$ being the wire length. 

In the absence of PSOI (\(v_{-} = 1\)), this expression turns into \( G_{\omega} = G_0 {\left(1 - i v_{+} \tan \frac{\omega \tau}{2 v_{+}}\right)}^{-1}\), in agreement with Ref.~\cite{PhysRevLett.81.1925}. The real part of the admittance $G'$ goes to zero at resonant frequencies \(\omega = \pi (2 n + 1) v_{+}\tau^{-1} \), which are multiple integer of the inverse transit time of the collective modes through the quantum wire. The only collective mode contributing to the electron current in this case is the plasmon excitation with the velocity \(v_{+} \), renormalized by the e-e interaction. Consequently, from zeros of \(G'\) one can extract information about the e-e interaction~\cite{PhysRevLett.81.1925,PhysRevB.58.13847}.

When PSOI is present, both collective modes contribute to the admittance with certain weights. The resulting oscillatory pattern that reflects the Fabry-Pérot resonances at the wire length has now two different characteristic frequencies, corresponding to different transit times of the slow and fast collective mode. The pattern is illustrated in Fig.~\ref{fig_5_4}. The mode velocities can be found directly from the measurements of the frequency dependence of the admittance because at sufficiently strong PSOI one of the velocities is strongly suppressed, which leads to a pronounced picture of double-periodic oscillations.

\section{Conclusion}
In the review, we presented a new view on the role of the spin-orbit interaction in solids associated with the presence of PSOI, and discussed the PSOI manifestations in some actual low-dimensional structures. The central theme is the appearance of an unexpectedly strong component of the pair interaction of electrons, which is determined by their spins and momenta, as well as by the strength and configuration of the Coulomb fields of the electrons. In crystals, the PSOI arises similarly to the SOI component of the Breit-Pauli Hamiltonian in the relativistic quantum theory. The important difference is that the PSOI component, caused by the magnetic field of moving electrons in the Breit theory, is extremely small in the crystals, while the component that is due to the Coulomb electric field of the electrons, in contrast, is very large similarly to the Rashba SOI\@. Accordingly, the PSOI appears to be important in materials with a strong Rashba SOI\@.

The PSOI has some remarkable features, the most interesting of which is the attraction of electrons that depends on the spins and momenta of the interacting electrons. The full range of possible effects that can occur due to the PSOI remains to be understood. But it is already clear that the attraction due to the PSOI leads to the formation of two-electron bound states. There are two pairing mechanisms that result in the formation of pairs of different types. The PSOI arising due to the relative motion of the electrons gives rise to the relative states, the properties of which do not depend on the motion of the pair as a whole. Also, for a definite spin configuration the PSOI creates an attraction that directly depends on the total momentum of a pair and disappears when the momentum is zero. In this way the convective states are formed when the total momentum is large enough. This is the most unusual effect of the PSOI\@. As far as we know, such a mechanism of the bound state formation has not yet been encountered in quantum mechanics.

Another feature of the bound states formed by the PSOI is that their binding energy as well as the spin and orbital structure in a nontrivial manner depend on the dielectric environment and on the presence of a metal gate. This is due to the fact that attraction is created not by the potential, but by the electric field, the configuration and strength of which change under the influence of the above factors. The effect of the dielectric screening and gates has been studied for a wide range of currently relevant structures: 1D wires with a gate, freely suspended 2D systems, 2D systems in a dielectric environment, and gated 2D layers. The dielectric screening decreases the binding energy, the stronger, the greater the polarizability of the layer or the wire. In addition the dielectric screening leads to a qualitative change in the electron density distribution. But of most interest is the gate effect, which is created by both image charges induced on the gate by the interacting electrons, and the voltage applied to it. In 1D wire, where the bound states arise only due to the image charges, the electric field produced by the gate voltage leads to a significant increase in the binding energy. In 2D systems, where the bound states are formed due to the in-plane Coulomb fields, the electric field of the gate changes them significantly, especially in the case when the ground state is degenerate at zero gate voltage. Thus, the gate allows one to effectively manage the electron pairs.

The binding energy of electron pairs reaches high values in materials with a sufficiently large Rashba SOI constant, which is nevertheless attainable at present. When the dielectric screening is minimal, the energy can be as high as tens of meV and, in principle, can be increased by the gate voltage. The high binding energy of pairs indicates the stability of the electron pairs and raises a very interesting question about the collective behavior of such pairs and, in general, the many-particle effects due to the PSOI in solids.

The presence and interplay of a long-range repulsion of electrons together with a more local attraction opens up a wide range of possible scenarios for the formation of a many-particle state, which includes no doubt the superconducting state and such phenomena as the formation of electronic complexes, the spontaneous reduction of the symmetry of the system and the clustering of electrons into geometric structures. A comprehensive study of various aspects of many-particle states is an interesting problem that requires further research.

To date, the many-electron problem was studied only for 1D systems with the use of various approaches limited to considering large-scale fluctuations. When the SOI is not very strong, the PSOI violates the spin-charge separation and changes the collective modes. Instead of pure charge and spin modes that exist without SOI, two modes with a mixed spin-charge structure are formed, one of which softens strongly in the long-wavelength region as the SOI grows. At critical value of the SOI parameter the system loses stability, which indicates that the system tends to radically transform its ground state.

PSOI is a new topic in the physics of electron systems in solids, which was formed quite recently and is still very little studied, and therefore there is a large number of unexplored problems. The fundamental difference between the PSOI and the widely studied Rashba SOI is that PSOI manifests itself directly in the interaction between the particles and, thus, significantly affects the structure of the correlated electronic state. PSOI has a universal nature, although its manifestations certainly depend on the details of a particular electron system.  There is no doubt that the PSOI also exists in topologically non-trivial electron systems, Dirac and Weyl semimetals, but the description of the PSOI in these systems may differ significantly from the simple single-band model with quadratic band spectrum considered here. To the best of our knowledge, such research has not been conducted in the literature yet, but it is evident that the most spectacular feature of the PSOI namely the short-ranged attraction of electrons in specific spin configurations paired with the long-ranged Coulomb repulsion should hold true for each of the systems above. Such unusual form of the e-e interaction could lead to the non-trivial effects in the topological superconducting systems built upon the 1D or 2D electron system with SOI and a superconductor even in the case when the PSOI is not particularly strong. Therefore, the concept of PSOI opens a promising direction of the future research of the strongly correlated electron systems.

\section*{Acknowledgments}
This work was carried out in the framework of the state task and partially supported by the Presidium of the Russian Academy of Sciences, Program No13 ``Fundamentals of high technology and the use of features of nanostructures in the natural science''.

\bibliography{review}

\end{document}